\documentclass[showpacs,superscriptaddress,aps,epsfig,psfig,rotate,rotating,pre]{revtex4}
\usepackage[final]{graphicx}
\begin{document}
\title{Critical Behavior of an Ising System on the Sierpinski Carpet: 
A Short-Time
Dynamics Study.}
\author{M. A. Bab}
\email{mbab@inifta.unlp.edu.ar} 
\affiliation{Instituto de Investigaciones Fisicoqu\'{i}micas Te\'{o}ricas y Aplicadas, Facultad de Ciencias Exactas,
Universidad Nacional de La Plata, Sucursal 4, CC 16 (1900) La Plata, Argentina.}
\affiliation{Departamento de F\'{i}sica, Facultad de Ciencias Exactas, Universidad Nacional de La Plata, 
La Plata, Argentina}

\author{G. Fabricius} 
\email{fabriciu@fisica.unlp.edu.ar}
\affiliation{Instituto de Investigaciones Fisicoqu\'{i}micas Te\'{o}ricas y Aplicadas, Facultad de Ciencias Exactas,
Universidad Nacional de La Plata, Sucursal 4, CC 16 (1900) La Plata, Argentina.}
\affiliation{Departamento de F\'{i}sica, Facultad de Ciencias Exactas, Universidad Nacional de La Plata, 
La Plata, Argentina}

\author{E. V. Albano}
\email{ealbano@inifta.unlp.edu.ar} 
\affiliation{Instituto de Investigaciones Fisicoqu\'{i}micas Te\'{o}ricas y Aplicadas, Facultad de Ciencias Exactas,
Universidad Nacional de La Plata, Sucursal 4, CC 16 (1900) La Plata, Argentina.}

\begin{abstract}
The short-time dynamic evolution of an Ising model embedded in an infinitely
ramified fractal structure with noninteger Hausdorff dimension was studied
using Monte Carlo simulations. Completely ordered and disordered spin
configurations were used as initial states for the dynamic simulations. In
both cases, the evolution of the physical observables follows a power-law
behavior. Based on this fact, the complete set of critical exponents
characteristic of a second-order phase transition was evaluated. Also, the
dynamic exponent $\theta $ of the critical initial increase in
magnetization, as well as the critical temperature, were computed. The
exponent $\theta $ exhibits a weak dependence on the initial (small)
magnetization. On the other hand, the dynamic exponent $z$ shows a
systematic decrease when the segmentation step is increased, i.e., when the
system size becomes larger. Our results suggest that the effective
noninteger dimension for the second-order phase transition is noticeably
smaller than the Hausdorff dimension. Even when the behavior of the
magnetization (in the case of the ordered initial state) and the
autocorrelation (in the case of the disordered initial state) with time are
very well fitted by power laws, the precision of our simulations allows us
to detect the presence of a soft oscillation of the same type in both
magnitudes that we attribute to the topological details of the generating
cell at any scale.
\end{abstract}

\maketitle

\section{Introduction.}

\smallskip In the last years, fractal structures with noninteger Hausdorff
dimension ($d_H$) have attracted the interest of researchers because these
systems, besides serving to model natural materials such as porous rocks,
aerogel, etc. \cite{buha,buha1}, also offer the possibility of theoretically
exploring systems exhibiting critical behavior close to their lower critical
dimension, i.e., the larger integer dimension in which the system does not
exhibit any phase transition at a finite temperature.

The first studies of phase transitions using fractal structures are those of
Gefen and co-workers \cite{1,2,3}. Based on renormalization methods, it has
been shown that a second-order phase transition at nonzero temperature
occurs only if the fractal substrate has an infinite ramification order.
Moreover, since translational symmetry is a necessary condition to proceed
with dimensional perturbation \cite{1}, the disagreement between the
critical exponents determined by current methods \cite{4,5,6,7,8,9} and
those obtained by continuation of $\varepsilon -$expansions to noninteger
dimension \cite{10} can be related to the topological features of the
fractal structure.

It is well known that for systems with translational symmetry, the influence
of the underlying structure becomes negligible at the critical point, i.e.,
when the correlation length

is much larger than the cell spacing, and only the dimensionality, the
number of components of the order parameter together with its symmetry and
the nature of the couplings concur to determine the values of the critical
exponents and the corresponding universality class. However, in fractal
systems, where the translational symmetry is replaced by scale invariance,
the topological details of the generating cell are present at any scale and
such universal behavior is said to be weak. The critical exponents and the
critical temperature depend not only on $d_H$, but also on the connectivity
and lacunarity of the fractal \cite{11}. A direct quantitative study of
topological effects has been recently published \cite{12}.

Most of the previously cited studies are based on the same type of fractal,
i.e., the Sierpi\'{n}ski Carpet (SC), which has an infinite ramification
order. Although the same kind of magnetic interaction (Ising model) has been
considered, these previous studies yield controversial results. Table \ref
{tab1} summarizes the list of published results for the case of the two
dimensional SC with $d_H=1.8927$, where the generating cell is built by
segmenting a square into nine subsquares and removing the central one, so
that this fractal is termed SC(3,1). In particular, different values have
been obtained by different authors for the critical temperature ($T_c$, see
table \ref{tab1}), which in most cases is lower than $T_{c}$ of the Ising
model in 2d. Also, a considerable scattering in the data corresponding to
the critical exponents $\beta/\nu$ and $\gamma/\nu$ can be observed. It
should also be noticed that Monte Carlo simulations and FSS may either
predict $d_{H} > d_{eff}$ \cite{5,7} or $d_{H} < d_{eff}$ \cite{6}.
\begin{table}[tbp]
\caption{List of critical temperatures and critical exponents reported in
the literature for the Ising model in the SC(3,1) fractal. The methods used
to obtain the data are: RSGR $\equiv$ Real Space Group Renormalization;
MC-FSS $\equiv$ Monte Carlo simulations and Finite-Size Scaling; MC-RNG $%
\equiv$ Monte Carlo Renormalization Group; MC-Slope $\equiv$ Monte Carlo
simulations and the slope method \protect\cite{8}. The index $k$ indicates
the generation of fractal used. The Boundary Conditions (BC) used are either
Periodic (P) or Free (F).}

\label{tab1}\centering
\begin{tabular}{ccccccccccc}
Ref. & $T_c$ & $\nu$ & $\frac \beta \nu $ & $\frac \gamma \nu $ & $d_{eff}$
& $k$ & BC & Method & Spins location &  \\ \hline\hline
\cite{4} & 2.06 & 1.12 & - & - & - & - & P & RSRG & Vertices &  \\ 
\cite{1} & 3.12 & - & - & - & - & - & F & RSRG & Vertices &  \\ 
\cite{6} & 1.482(15) & 1.565(10) & 0.0815(30) & 1.76(1) & 1.923(16) & 7 & P
& MC-FSS & Center &  \\ 
\cite{6} & 1.482 & 1.73(3) & 0.147(9) & 1.625(20) & 1.919(28) & 7 & F & 
MC-FSS & center &  \\ 
\cite{5} & 1.481 & 1.70(1) & 0.080(1) & 1.730(1) & 1.890(2) & 7 & P & MC-FSS
& Center &  \\ 
\cite{7} & 1.4795(5) & - & 0.075(10) & 1.732(4) & 1.882(24) & 8 & P & MC-FSS
& Center &  \\ 
\cite{8} & 1.4992(11) & - & - & - & $\simeq 1.7$ & 6 & P & MC-Slope & Center
&  \\ 
\cite{5} & 1.479546(16) & - & - & - & - & - & P & MC-RNG & Center &  \\ 
Ising & 2.269 & 1 & 0.125 & 1.75 & 2 & - & - & Exact & - &\\ \hline\hline 
\end{tabular}
\end{table}

One of the reasons explaining the discrepancies reported in the calculation
of $T_{c}$ is most likely related to the location of the spins on the
fractal (see Table \ref{tab1}), which could be at the vertices \cite{1,3,4,9}
or at the center \cite{5,6,7,8} of the squares. Consequently the mean number
of nearest neighbors per site is not the same. In addition, for the former
the number of spins as a function of the cell size does not follow a power
law and consequently, $d_H$ should not be expected to enter in the
description of the critical phenomena of such systems. Another reason
causing discrepancies is related to finite-size effects, since scaling
analyses reveal very strong scaling corrections for dimensions smaller than $%
d = 2$ \cite{5,6,7}. Monceau et al \cite{7} have noticed that discrepancies
with standard finite-size scaling (FSS) methods could also be due not only
to the operation of pure (genuine) finite-size effects, but also to a
topological contribution to scaling corrections \cite{7}. They have also
pointed out that the hyperscaling law $d_{eff}= 2\beta /\nu + \gamma/\nu $
should remain valid for $d_{eff} = d_H$ and the susceptibility should follow
the expected power-law behavior, which allows to calculating the ratio of
exponents $\gamma /\nu $ and the anomalous dimension exponent $\eta $ in a
reliable way \cite{13}. In addition, these authors have shown the agreement
of their results with those determined by Monte Carlo renormalization group
techniques \cite{14}.

On the order hand, Pruessner et al \cite{8} have questioned the validity of
FSS studies on fractal structures. In fact, they have pointed out that each
segmentation step represents a new thermodynamic system and cannot be
treated as a scaled version of the previous one. In order to avoid this
shortcoming, the authors have proposed the Slope Method \cite{8}. The
critical exponents obtained using this method suggest that $d_{eff}$ is
smaller than $d_H$ \cite{8}, in agreement with the 4-$\varepsilon $
renormalization group prediction. Furthermore, the obtained critical 
temperature is higher than those reported by other authors by using FSS
(see Table \ref{tab1}).

Another useful approach to obtain independent estimations for the critical
temperature and the critical exponents is given by the Short-Time Dynamics
(STD) \cite{18} method (for available results see rows $1-3$
of Table \ref{tab2}). By performing simulations using both the 4th and
5th segmentation steps, Pruessner et al \cite{8} have reported substantial
differences between the critical temperature estimated by the STD method
(see Table \ref{tab2}) and the data corresponding to  
the FSS approach (see Table \ref{tab1}). Furthermore, the 
exponents $\nu z$, $\gamma $, and $\beta $ (see also
table \ref{tab2}) strongly depend on the segmentation step. In addition, 
Zheng et al. \cite{9} have determined the exponent $\theta $ of the
initial increase in magnetization which seems to be
slightly greater than the figure accepted for the 2d Ising model. However,
this determination has to be taken with caution because spins are located at
the vertices of the fractal and this approach is expected to give an
inaccurate estimation of $T_c$, as has already been discussed.
\begin{table}[tbp]
\caption{List of critical temperatures and critical exponents reported in
the literature for the Ising model in the SC(3,1) fractal and obtained using
Monte Carlo simulations and the Short-Time Dynamics analysis (MC-STD). PW $%
\equiv$ Present Work. The Boundary Conditions used are Periodic and the
index $k$ indicates the generation of the fractal used.}
\label{tab2}\centering
\begin{tabular}{cccccccccc}
Ref. & $T_c$ & $\nu z$ & $\gamma $ & $\beta $ & $\theta $ & $z$ & $k$ & 
Spins location &  \\ \hline\hline
\cite{9} & 2.033(4) & - & - & - & 0.211(3) & 2.38(4) & 7 & Vertices &  \\ 
\cite{8} & 1.5266(11) & 3.06(11) & 1.959(32) & 0.1154(29) & - & - & 4 & 
Center &  \\ 
\cite{8} & 1.5081(12) & 3.21(15) & 2.048(49) & 0.120(55) & - & - & 5 & Center
&  \\ 
PW & 1.4945(50) & 3.546(12) & 2.22(1) & 0.121(5) & 0.1815(6) & 2.55(1) & 6 & 
Center &  \\ 
Ising & 2.269 & 2.165\cite{23} & 1.75 & 0.125 & 0.191\cite{23} & 2.165\cite
{23} & - & - & \\ \hline\hline
\end{tabular}
\end{table}

In view of the scattering of the available data for the critical temperature
and critical exponents, the aim of this paper is to study the critical
behavior of the Ising Model on the fractal structure SC(3,1) using the
Short-Time Dynamics approach. In order to achieve this goal, we have used a
segmentation step bigger than in a previous STD study of this system \cite{8}%
. Furthermore, we have confirmed that the time dependence of the
magnetization follows a power-law behavior for times two orders of magnitude
larger than in a previous STD study \cite{8}. Additionally, we have
determined $T_c$ and the complete set of critical exponents starting the STD
studies with two different initial conditions: i) a fully ordered initial
state (ground state configuration corresponding to $T = 0$), and ii) a fully
disordered initial state corresponding to $T = \infty$. Self-consistency of
the results was also carefully checked. On the other hand, we have also
obtained the value of the exponent $\theta$ related to the initial increase
in magnetization at an early time, which have not been previously determined
for spins placed at the center of the occupied subsquares of the fractal.
Finally, the dynamic exponent $z$ was evaluated, for first time for this fractal, 
by means of two independent methods using the Binder's cumulant and the critical
scaling of time correlation functions.     

The STD approach has shown to be a powerful tool in the study of critical
phenomena because the critical exponents can be determined before the
critical slowing down of the dynamics takes place and they are free from
finite size corrections, provided that the correlation length $\xi(t)$ 
is always smaller than the system size $L$ ($\xi(t) \ll L$) \cite{15,16}. 
However, in the case that we are interested in, namely the SC(3,1), 
an intrinsic kind of finite-size effects due to the segmentation step $k$ 
of the fractal can be observed, as it is discussed in details below. 
Furthermore, the STD analysis is particularly useful
since the application of FSS to obtain $T_c$ is questionable for
segmentation steps smaller than $6$ \cite{8}, which are
the typical sample sizes that one is forced to use in practical
calculations due to computing limitations \cite{5,6,7}.

The outline of this paper is the following: In Section II we describe the
magnetic model and the underlying fractal structure, in Section III we
briefly recall the main features of the Short-Time Dynamics method, and in
Section IV, the results obtained from the dynamic simulations are presented. 
Finally, our conclusions are stated and discussed in Section V.

\section{The Ising Model on the Sierpi\'{n}ski Carpet.}

\begin{figure}[ht]
\includegraphics[height=10cm,width=7cm,angle=-90]{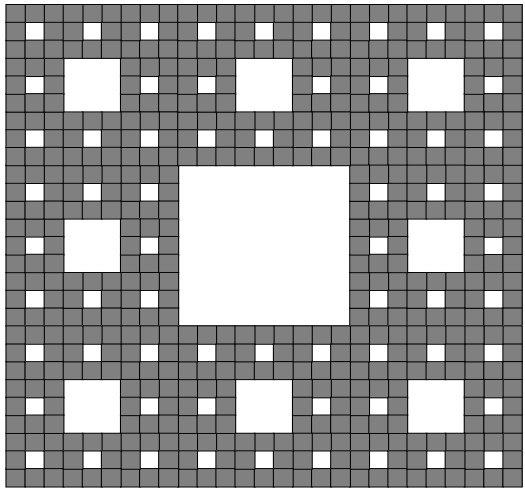} 
\caption{Sketch of the Sierpinski carpet SC(3,1) iterated up to $k=3$ 
segmentation step. Spins are placed at the center of the filled squares.}
\label{fig0}
\end{figure}

The SC(b,c) is obtained as follows: for each segmentation step ($k$), a
square of length $L$ is segmented into $b^2$ subsquares and $c^2$ subsquares
are deleted from the center of the initial square; then the segmentation
process is iterated on the remaining subsquares.  Figure \ref{fig0} shows a
sketch of the SC(3,1), which is used in the present work, corresponding 
to the $k = 3$ segmentation step. In the limit $k \rightarrow
\infty$ the mathematical fractal SC(b,c) is obtained, and the Hausdorff
dimension is given by $d_H=\frac{\ln (b^2-c^2)}{\ln b}$. In the case of the
SC(3,1) the deviation of the mean number
of near nearest neighbors from that corresponding to the thermodynamic limit
(mathematical fractal), using periodic boundary conditions, as determined by
the transfer-matrix method, becomes negligible for k $\geq $ 6\cite{7}.

As mentioned above, this fractal has an infinite ramification order, which
implies that the Ising model should exhibit a second-order phase transition
at finite temperature. Spins were placed at the center of the occupied
subsquares. Consequently the number of spins increases as a power law of the
lattice size, and the exponent is given by $d_H$.

The Hamiltonian of the system is given by

\begin{equation}
H=-J \sum _{\left\langle i,j\right\rangle } s_i\;s_j  \label{ec1}
\end{equation}

\noindent where $s_i$ assumes the values $\pm 1$, the sum runs over all
interacting nearest-neighbor pairs of spins, and the exchange coupling
constant $J$ is positive (ferromagnetic interactions).

\section{Short-Time Dynamics approach for critical phenomena.}

According to field theoretical calculations \cite{18}, if a magnetic system
at high temperature, with a small magnetization m$_0$, is suddenly quenched
to the critical temperature, it may exhibit a universal dynamic evolution,
which sets right after a time scale $t_{mic}$. It is expected that $t_{mic}$
should be large in the microscopic sense, but still very small in the
macroscopic sense necessary for equilibration. This STD approach is free of
the critical slowing down since the spatial correlation length is still
small within the short-time regime, even at (or near) the critical point 
\cite{18}.

The $k$-th moments of the magnetization are given by \cite{15,16}

\begin{equation}
M^k(t,\tau ,L,m_0)=b^{-\frac{k\beta }\nu }M^k(b^{-z}t,b^{\frac 1\nu }\tau
,b^{-1}L,b^{x_0}m_0)  \label{ec2}
\end{equation}

\noindent where $\beta$ and $\nu$ are the order parameter and the
correlation length critical exponents, $z$ is the dynamic critical exponent, 
$\tau $ is the reduced temperature, $L$ is the system size, $x_0$ is the
scaling dimension of the initial magnetization, and $b$ is a scaling factor.
For large $L$, at the critical point $\tau = 0$, and for $m_0\ll 1$, from
the scaling form given by equation (\ref{ec2}) one derives the initial
increase in magnetization, obtaining \cite{15,16}

\begin{equation}
M(t)=\left[ \left\langle \frac 1N\sum_{i=1}^Ns_i\right\rangle \right]
=m_0\;t^\theta F\left( t^{\theta +\frac \beta {\nu z}}\,m_0\right)
\label{ec3}
\end{equation}

\noindent where $\left\langle ...\right\rangle $ denotes the averages taken
over spin configurations and $\left[ ...\right] $ corresponds to averages
taken over different samples with equivalent initial conditions. Here, $%
\theta =\left( x_0-\frac \beta \nu \right) /z$, and the scaling function
behaves as $F(x) \sim 1$ for $x \rightarrow 0$ and $F(x) \sim \frac 1x$ for $%
x \rightarrow \infty $. It should be noticed that the time scale for this
initial increase is of the order of $t_{0}\sim m_{0}^{-\frac{z}{x_{0}}}$. $%
\theta$ and $x_{0}$ are the exponents of the initial increase and the
scaling dimension of the order parameter. Since both exponents are related,
one of them can be considered as a new {\it no trivial} critical exponent 
\cite{15,16}. Performing simulations for different values of the initial
magnetization and extrapolating the results to $m_0 = 0$, the exponent $%
\theta $ can be obtained. Other interesting observables are the second
moment of the magnetization ($M^{2}(t)$) and the autocorrelation ($A(t)$),
that for $m_0=0$ and $\tau =0$ should behave according to the following
power-law scaling relationships

\begin{equation}
M^2(t)=\left[ \left\langle \left( \frac 1N\sum_{i=1}^Ns_i\right)
^2\right\rangle \right] \propto t^{\left[ \frac{d_{eff}}z-\frac{2\beta }{\nu
z}\right]
},\,\,\,\,\,\,\,\,\,\,\,\,\,\,\,\,\,\,\,\,\,\,\,\,\,\,\,\,\,\,\,\,\,\,\,\,
\label{ec4}
\end{equation}

\noindent and

\begin{equation}
A(t)=\left[ \left\langle \frac 1N\sum_{i=1}^Ns_i(t)\,\,s_i(0)\right\rangle
\right] \propto t^{-\lambda},\,\, {\text with}\,\,\, \lambda = \frac{d_{eff}}%
z-\theta ,  \label{ec5}
\end{equation}

\noindent respectively.

Another important process that can be measured is the dynamic relaxation
from a completely ordered state (with $m_0=1$), which corresponds to a
ground state configuration at $T=0$, to $T_c$. In this case, the
magnetization, the logarithmic derivative of the magnetization with respect
to $\tau$, and the second-order Binder's cumulant \cite{23}
should behave according to

\begin{equation}
M(t) \propto t^{-\frac \beta {\nu z}} ,  \label{ec6}
\end{equation}

\begin{equation}
V_\tau (t)=\partial _\tau (ln\;M(t,\tau ))\mid _{\tau =0}\propto t^{\frac 1 {%
\nu z}} ,  \label{ec7}
\end{equation}

\noindent and

\begin{equation}
U(t)=\frac{M^2(t)}{\left[M(t)\right]^2} - 1\propto t^{\frac{d_{eff}}z} ,  
\label{ec8}
\end{equation}

\noindent  respectively. For $T\neq T_c$, but within the critical region,
the power-law behavior is modified by a scaling function, which for the
magnetization is given by $M(t^{\frac 1{\nu z}}\tau )$. This fact can be
used to determine the critical temperature from the localization of the
optimal power-law behavior.

Summing up, the STD scaling study of a given system performed by starting
from two extreme initial states, i.e. a completely ordered one and a
completely disordered one, is sufficient to determine both the critical
temperature and the set of relevant critical exponents in a self-consistent
fashion \cite{15,16,17}.

\section{Numerical Results.}

\subsection{Details on the Simulations.}

Monte Carlo simulations of the Ising model on the SC(3,1) were performed for
the segmentation step $k=6$ ($L = 729$ with 262144 spin sites) using
periodic boundary conditions, and starting either from an ordered state or
from a disordered state with zero or a small initial magnetization. In the
latter case the initial magnetization ($m_{0}$) was obtained from a
disordered configuration (of zero magnetization) by flipping a definite
number of spins at randomly chosen sites.

In order to implement the time evolution, the system is updated by using the
Metropolis algorithm. The well tested \cite{random} Marsaglia-Zanan 
pseudo random number generator is used throughout the simulations.
The time unit, defined as a Monte Carlo time step
(MCS), involves the update of a number of spins that corresponds to all spin
sites of the sample. In this way, during one MCS each spin is updated once,
on average. Simulations starting from a disordered (ordered) state are
carried out up to 2000 MCS (200000 MCS).

The magnetization, the autocorrelation, and the second moment of the
magnetization were averaged over a number $n_s$ of samples with equivalent
initial configurations. In addition, the time evolution of the ordered state
was also studied for the segmentation steps $k=3,4,5$ in order to apply an
FSS method that allowed us to obtain an estimation of the dynamic exponent $z$.
In order to estimate the error bars of the evaluated exponents we have used a 
variant of the blocking method \cite{error} fitting 
the time dependence 
of each observable for independent sets of measurements having the 
same statistic.

\subsection{Simulation of the Dynamic Evolution starting from the ordered
state.}

\begin{figure}[ht]
\includegraphics[height=10cm,width=8cm,angle=-90]{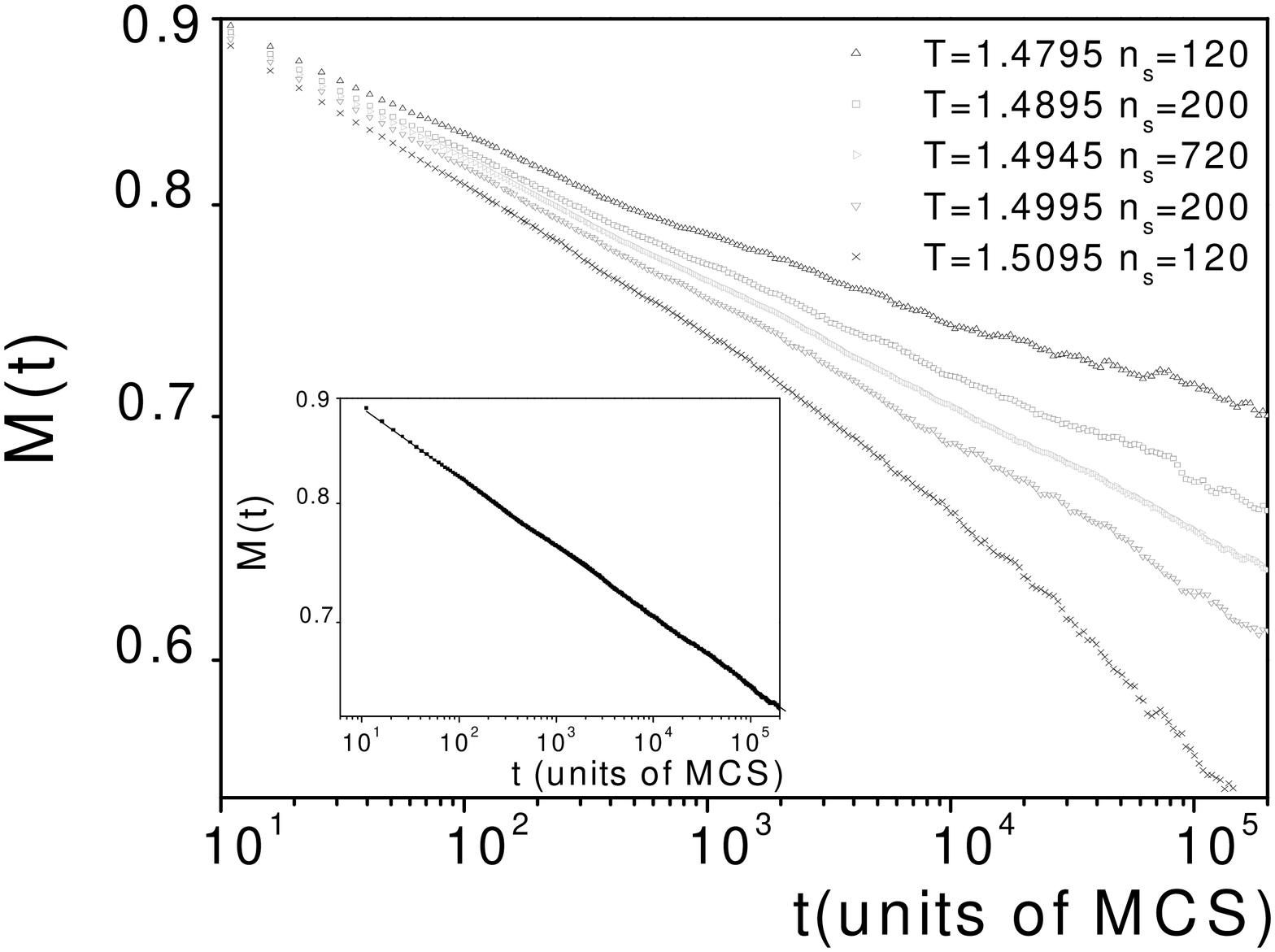} 
\vskip 1.0 true cm
\caption{Log-log plots of magnetization versus time obtained 
starting from ordered initial conditions ($m_{0} = 1)$. The different
temperatures used and the number of samples considered are also 
shown in the figure. The inset shows the log-log plot
of $M(t)$ versus $t$ that gives the best fit of equation (\ref{ec6})
and is assumed to correspond to the critical temperature $T_{c} = 1.4945$.}
\label{fig1}
\end{figure}

According to our experience, the determination of the critical temperature
and exponents is more accurate when the simulations start from the ordered
state. In fact, in this case the magnetization is large and decreases slowly
during time evolution and therefore statistical fluctuations are less
prominent. Figure \ref{fig1} shows the decay of the magnetization obtained
at different temperatures for $k=6$. The critical temperature is determined
by finding the smallest standard deviation from the power law given by
equation (\ref{ec6}), which yields $T_c(k=6)=1.4945(50)$, where the error
bar is assessed by considering the closest pair of temperatures that present
noticeable but small standard deviations.
This is in good agreement with determinations performed by means of 
Monte Carlo simulations analyzed by using Slope method \cite{8}. Also, 
acceptable agreement with the value obtained by means of the STD analysis of
Monte Carlo data is found (see Table \ref{tab2}). However, a careful 
inspection of the data shows a systematic decrease in $T_{c}$ when the 
segmentation step is increased, suggesting that our result could be taken as
an upper bound.

The exponent determined by fitting
equation (\ref{ec6}), see the inset of Figure \ref{fig1}, is listed in Table 
\ref{tab4} (2nd column). Our estimation, given by $\beta /\nu z=0.0341(1)$,
is slightly smaller than the figures obtained by using the exponents listed
in Table \ref{tab2}, which correspond to STD studies for $k=4$ and $5$,
namely $\beta /\nu z=0.0377$ and $\beta /\nu z=0.0374$ \cite{8},
respectively. So, our finding is consistent with a systematic decrease in $%
\beta /\nu z$ that is observed when $k$ is increased.

\begin{table}[tbp]
\caption{List of critical exponents determined from the time dependence of
the magnetization (second column, see equation (\ref{ec6})), the Binder's
cumulant (third column, see equation (\ref{ec8})), the logarithmic
derivative of magnetization (fourth column, see equation (\ref{ec7})), and
the susceptibility (fifth column). Data obtained starting the simulations
from an ordered initial state and for the k= 6 generation of the fractal.
Slightly different exponents are obtained performing the fits after
disregarding different initial time intervals (${\bf t}_{min}$), as listed
in the first column.}
\label{tab4}\centering
\vskip 2.0 true cm 
\begin{tabular}{ccccccc}
${\bf t}_{min}{\bf (MCS)}$ & $\frac \beta {\nu z}$ & $\frac{d_{eff}}z$ & $%
\frac 1{\nu z}$ & $\frac \gamma {\nu z}$ & ${\bf \beta }$ & ${\bf \gamma }$
\\ \hline\hline
20 & 0.03406(6) & 0.697(3) & 0.285(2) & 0.630(3) & 0.119(9) & 2.21(2) \\ 
100 & 0.03412(7) & 0.693(2) & 0.282(1) & 0.626(2) & 0.121(5) & 2.22(1) \\ 
150 & 0.03413(7) & 0.694(2) & 0.282(2) & 0.627(2) & 0.121(9) & 2.22(2)\\ \hline\hline
\end{tabular}
\end{table}

\begin{figure}[ht]
\includegraphics[height=10cm,width=8cm,angle=-90]{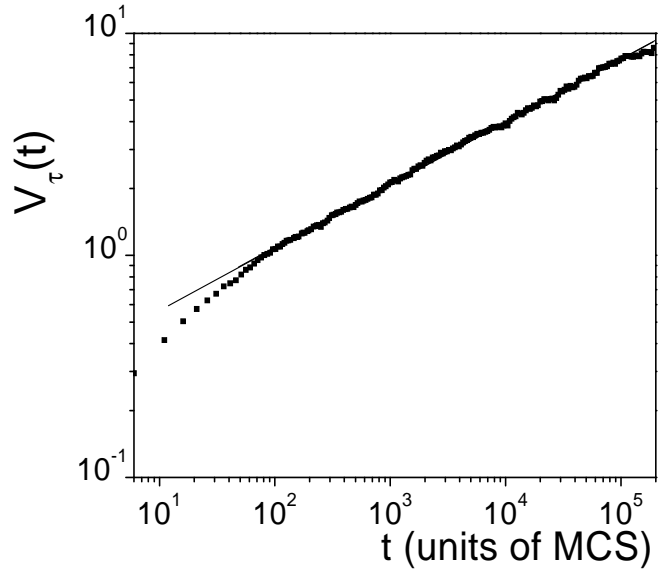} 
\vskip 1.0 true cm
\caption{Log-log plot of the logarithmic derivative of magnetization 
versus time obtained at criticality, starting from ordered 
initial conditions ($m_{0} = 1)$. The full line corresponds to
the best fit obtained for $t > 100$ MCS, 
according to equation (\ref{ec7}).}
\label{fig2}
\end{figure}

The logarithmic derivative of the magnetization with respect to $\tau$ is
evaluated by taking the difference between the values of $M(t)$ at two
temperatures close to $T_c$. The result of this calculation, obtained taking 
$T_1 = 1.4795$ and $T_2 = 1.5095$, is shown in Figure \ref{fig2}. From this
figure it follows that the power-law behavior expected from equation (\ref
{ec7}) is obtained after $t=70MCS$, and the corresponding exponent is listed
in Table \ref{tab4}. It should be noticed that using $T_1 = 1.4895$ and $T_2
= 1.4995$ one also obtains the same critical exponent (within error bars)
but the data are more noisy. From the data shown in Table \ref{tab4} it
follows that $\nu z = 3.546$, which is significantly larger than the
exponent reported for $k = 5$, namely $\nu z = 3.21$ \cite{8}.

\begin{figure}[ht]
\includegraphics[height=10cm,width=8cm,angle=-90]{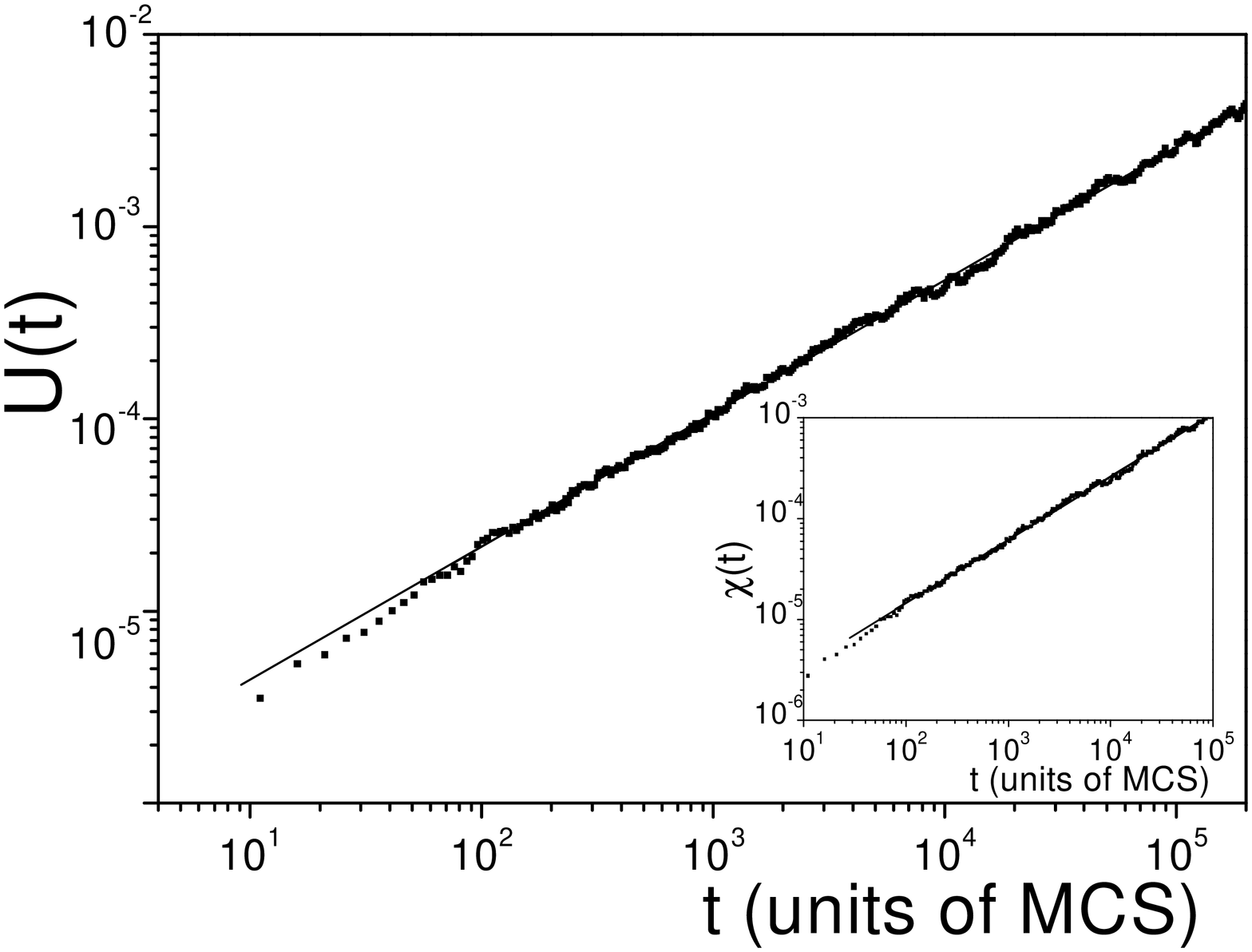} 
\vskip 1.0 true cm
\caption{Log-log plot of Binder's cumulant versus time obtained at criticality, 
starting from ordered initial conditions ($m_{0} = 1)$. The inset shows the
zero field susceptibility ($\chi$) versus time. 
The full lines correspond to 
the best fits obtained for $t > 100$ MCS.}
\label{fig3}
\end{figure}

Figure \ref{fig3} shows the determination of the exponent $\frac{d_{eff}}z$ (see the
results in Table \ref{tab4}) from the time dependence of the Binder
cumulant, according to equation (\ref{ec8}). Considering that the
susceptibility is given by $\chi (t)\propto U(t)\left[ \left\langle
M(t)\right\rangle \right] ^2\propto t^{\frac \gamma {\nu z}}$, the exponent $%
\gamma/\nu z $ can also be obtained (see inset in Figure \ref{fig3} and Table \ref
{tab4}). Notice that the above relationships were obtained assuming that the
hyperscaling law $d_{eff}= 2\beta /\nu + \gamma/\nu $ holds.

Simulations started from the ordered stated also allow the self-consistent
determination of the order parameter critical exponent $\beta$ and the
exponent $\gamma$ of the susceptibility using the exponents listed in the
2nd and 4th columns, and in the 5th and 4th columns of Table \ref{tab4},
respectively. The obtained results are also listed in Table \ref{tab4}. The
trend of the data for the exponent $\gamma$, namely a systematic increase
with $k$, is consistent with the observations reported by Pruessner et al. 
\cite{8}, which are listed in Table \ref{tab2} for the sake of comparison.
However, $\beta$ appears to be less sensitive to the change of the
segmentation step. It is worth mentioning that our best estimation of the
order parameter critical exponent given by $\beta = 0.121(5)$ is very close
to the exact value corresponding to the Ising model in $d = 2$ ($\beta =
0.125$). However, $\gamma$ is clearly greater in the case of the fractal
substrate.

\subsection{Simulations of the Dynamic Evolution starting from a disordered
state}

\begin{figure}[ht]
\includegraphics[height=10cm,width=8cm,angle=-90]{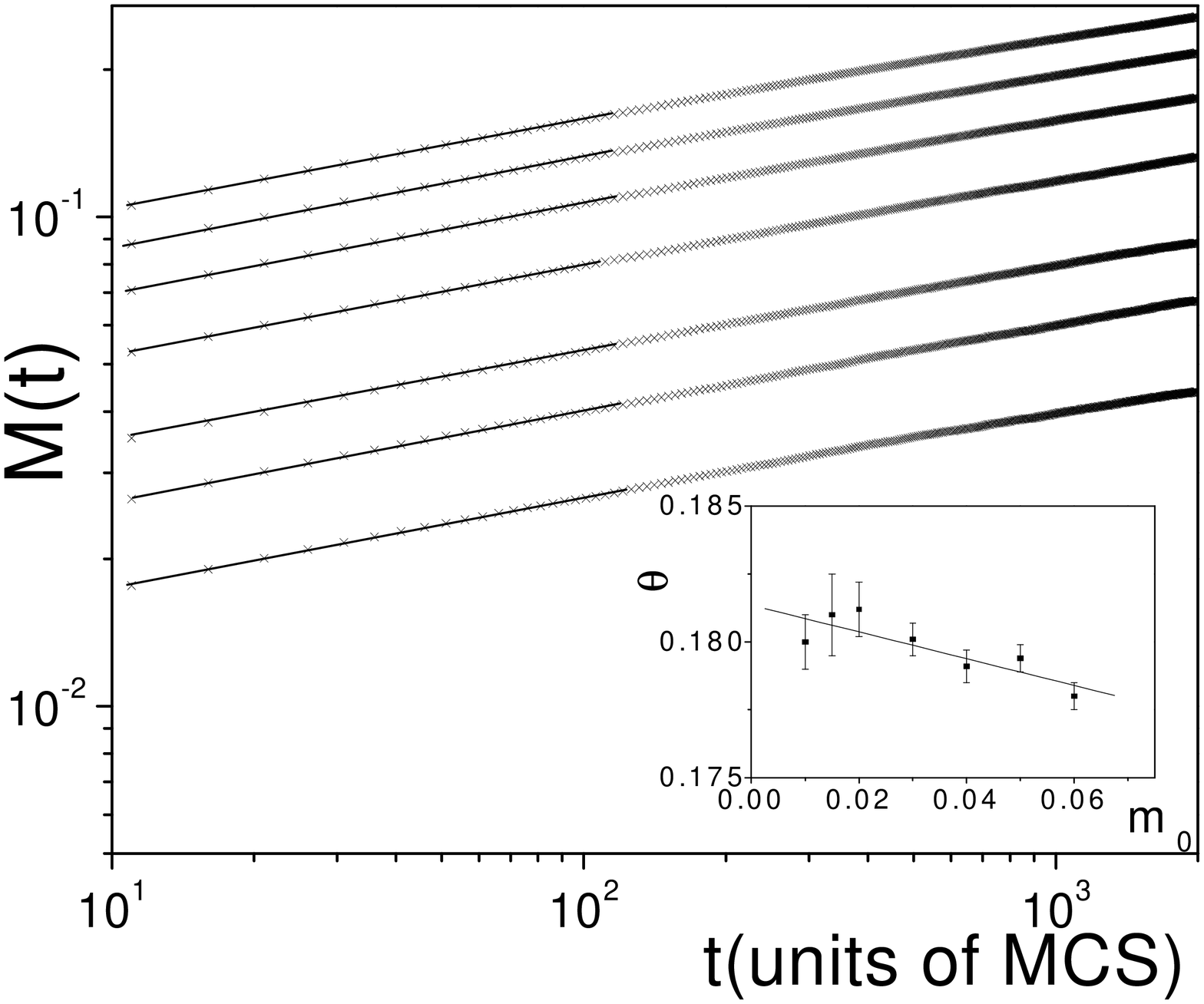} 
\vskip 1.0 true cm
\caption{Log-log plots of magnetization versus time obtained at criticality, 
starting from disordered initial conditions slightly modified 
to obtain different values of the initial magnetization $m_{0}$. 
Data corresponding to $k = 6$ and different values of $m_{0}$, 
which from top to bottom are: $0.06, 0.05, 0.04, 0.03, 0.02, 0.015$ and $0.01$,
respectively. The full lines correspond to the 
best fits obtained for $20 \le t \leq 100$ MCS, 
according to equation (\ref{ec3}). 
Data obtained by averaging over $4000-10000$ different samples,
depending of $m_{0}$. The inset shows the dependence 
of $\theta$ on the  initial magnetization $m_{0}$ that allowed us to 
extrapolate the exponent $\theta(m_{0} \rightarrow 0) = 0.1815(6)$.
More details in the text.}
\label{fig4}
\end{figure}

Figure \ref{fig4} shows the initial increase in magnetization, observed for different
values of the initial (small) magnetization ($m_{0}$), and obtained after
quenching the system to $T_c$ when the simulations started from the
disordered state corresponding to $T = \infty$.
Within the time regime considered ($20-2000$ MCS), the magnetization always
increases and the data can be fitted to a power law with critical exponent $%
\theta$, as expected from equation (\ref{ec3}). Nevertheless, a soft
curvature of the data can be observed for larger times due to the fact that $%
m_{0}$ is finite and the power law is actually expected to hold in the $%
m_{0} \rightarrow 0$ limit. So, in order to determine the critical exponent
we performed a fit of the data within the time interval 20-100MCS. As can be
observed in the inset of figure \ref{fig4}, the exponents show a weak dependence on $%
m_0$. Then, the exponent $\theta$ was evaluated by a linear extrapolation to 
$m_{0}=0 $, yielding $\theta = 0.1815(6)$. So, according to our results, the
exponent $\theta$ for the SC(3,1) fractal appears to be slightly smaller
than the accepted value for the Ising model in $d = 2$, given by $\theta =
0.191$ \cite{peter}.

\begin{figure}[ht]
\includegraphics[height=10cm,width=8cm,angle=-90]{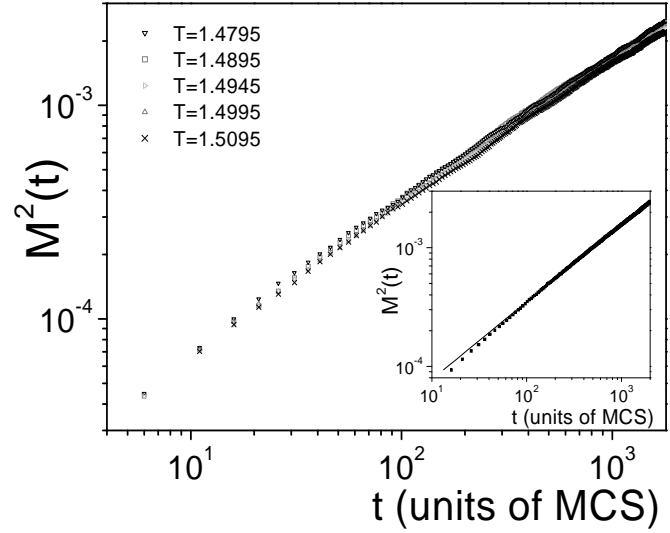} 
\vskip 1.0 true cm
\caption{Log-log plots of the second moment of magnetization versus time,
obtained at different temperatures (listed in the figure), 
starting from disordered initial conditions  with $m_{0} = 0$. 
Data obtained by taking $k = 6$.
The inset shows the dependence of $M^{2}$ on time obtained
at criticality ($T_{c} = 1.4945$), where the exponent  
$\frac{d_{eff}}z{\bf -}\frac{2\beta } {\nu z}$ is obtained by fitting 
the data according to equation (\ref{ec4}). See also Table \ref{tab5}.
Data obtained by averaging over 8000 different samples.}
\label{fig5}
\end{figure}

Figure \ref{fig5} shows that the evolution from a disordered state, with $m_0=0$, of
the second moment of the magnetization has a weak dependence on temperature.
This shortcoming hinders an independent estimation of $T_{c}$ based on these
measurements. However, by using the value of $T_{c}$ obtained by means of
simulations started from the ordered state, it is possible to evaluate the
critical exponent of the second moment according to equation (\ref{ec4}), as
shown in the inset of figure \ref{fig5}. The obtained value is listed in Table 
\ref{tab5}.

\begin{table}[tbp]
\caption{List of critical exponents determined from the dynamic behavior of
the second moment of the magnetization (second column, see equation (\ref
{ec4})), and autocorrelation (third column, see equation (\ref{ec5})). Data
obtained starting the simulations from disordered initial states with $m_{0}
= 0$ and for $k = 6$. The estimations of $d_{eff}/z$ listed in columns 4th
and 5th are obtained by using the autocorrelation data (3rd column) and the
determined value of $\theta = 0.1815$, and using the exponents listed in the
2nd column in combination wich the exponents listed in the 2nd column of
Table \ref{tab4}, respectively. }
\label{tab5}\centering
\vskip 2.0 true cm 
\begin{tabular}{ccccc}
${\bf t}_{min}{\bf (MCS)}$ & $\frac{d_{eff}}z{\bf -}\frac{2\beta } {\nu z}$
& $\frac{d_{eff}}z{\bf -\theta }$ & $\frac{d_{eff}}z$ (*) & $\frac{d_{eff}}z$
(**) \\ \hline\hline
20 & 0.665(2) & 0.518(2) & 0.699(2) & 0.699(2) \\ 
100 & 0.648(2) & 0.514(2) & 0.695(2) & 0.682(2) \\ 
150 & 0.646(2) & 0.512(3) & 0.693(3) & 0.680(2)\\ \hline\hline
\end{tabular}
\end{table}

\begin{figure}[ht]
\includegraphics[height=10cm,width=8cm,angle=-90]{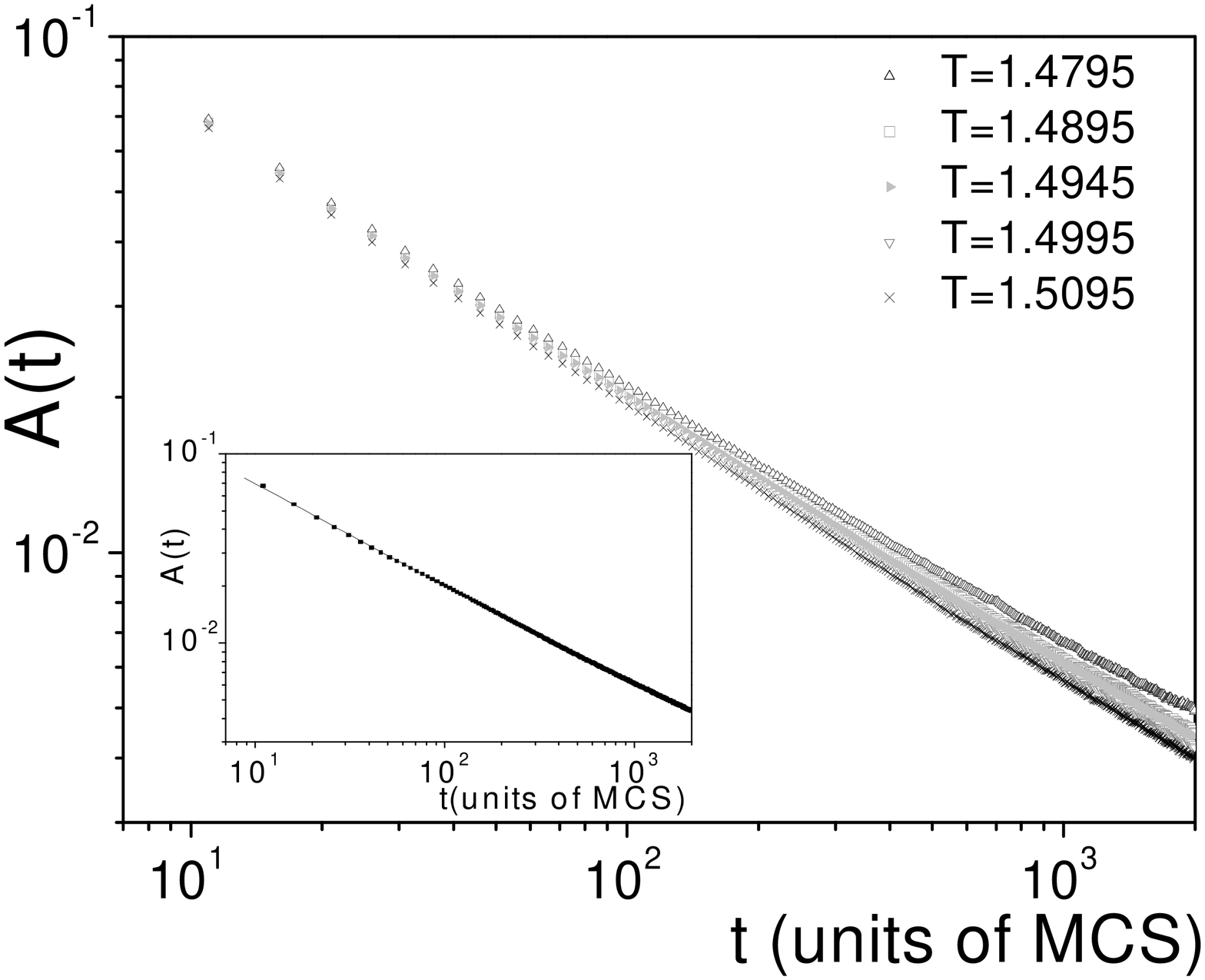} 
\vskip 1.0 true cm
\caption{Log-log plots of autocorrelation versus time,
obtained at different temperatures (listed in the figure), 
starting from disordered initial conditions  with $m_{0} = 0$
with $k = 6$. The inset shows the dependence of 
autocorrelation on time obtained
at criticality ($T_{c} = 1.4945$), where the exponent  
$\lambda = \frac{d_{eff}}z{\bf -}\theta $ is obtained by fitting 
the data according to equation (\ref{ec5}). See also Table \ref{tab5}.
Data obtained by averaging over 8000 different samples.}
\label{fig6}
\end{figure}

On the other hand, the decay of the autocorrelation function (see figure \ref{fig6})
slightly depends on $T$, allowing us to confirm our estimation, namely 
$T_{c} = 1.4945(50)$, and the corresponding error bars, already evaluated
using simulations started from ordered configurations. The exponent $\lambda
= d_{eff}/z - \theta$ evaluated by fitting the data at criticality (see
inset of figure \ref{fig6}) is also listed in Table \ref{tab5}. It is worth
mentioning that by inserting the value of $\theta$ already determined in the
exponent of the autocorrelation function, one can also calculate $d_{eff}/z$%
, as listed in the 4th column of Table \ref{tab5}. The obtained results are
in full agreement with the determination performed by starting simulations
from ordered states, see Table \ref{tab4}.

Also, inserting the exponent $\beta /\nu z$ determined by means of
simulations started from the ordered state (see the 2nd column of Table 
\ref{tab4}) in the expression of the exponent of the second moment of the
magnetization, given by $d_{eff}/z-2\beta /\nu z$, one can obtain an
additional estimation of $d_{eff}/z$, as listed in the 5th column of Table 
\ref{tab5}, which is slightly smaller than the estimations already performed
by means of different procedures. We attribute this small difference to the 
propagation of errors in the evaluation of exponents by combining results 
from different measurements.

\subsection{Determination of the dynamic exponent $z$.}

\begin{figure}[ht]
\includegraphics[height=10cm,width=8cm,angle=-90]{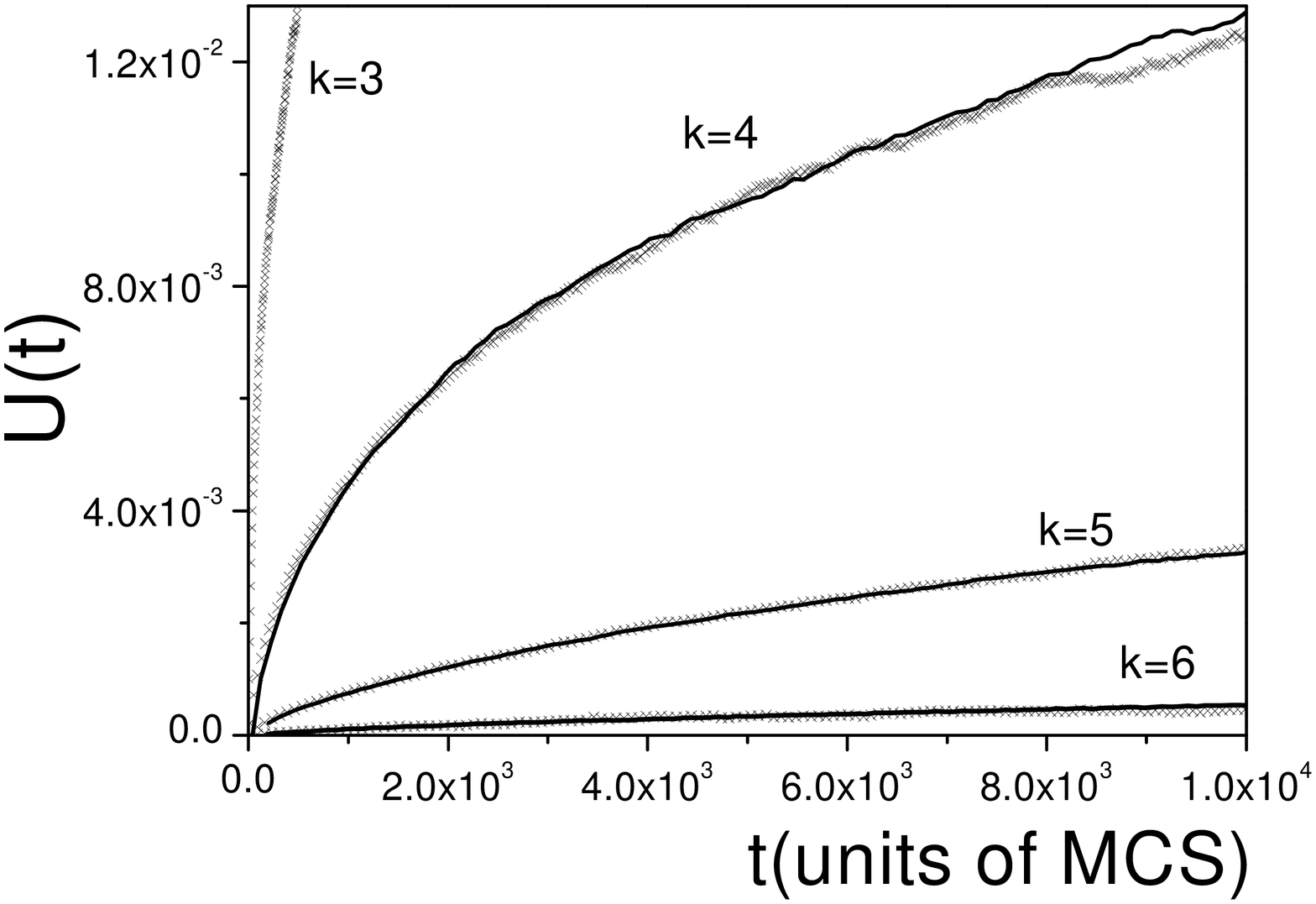} 
\vskip 1.0 true cm
\caption{Scaling plots of the second-order Binder's cumulant obtained 
using data corresponding
to adjacent pairs of segmentation steps ($k_{i}, k_{i+1}$) with sizes
($L_{i}, L_{i+1}$), respectively. The crosses (solid lines) correspond to 
systems of sizes $L_{i+1}$ ($L_{i}$). The time scale for the latter is
rescaled by a factor $(L_{i+1}/L_{i})^{z}$ in order to achieve superposition
to the former; in each case the exponent $z$ is taken as a
fitting parameter.
Data obtained by averaging over $720, 17000, 47000$ and $70000$ different 
samples, for $k = 6, 5, 4$ and $3$, respectively.}
\label{fig7}
\end{figure}

Although the application of FSS techniques to the evaluation of critical
exponents in systems with fractal structure is questionable, as we have
already mentioned in the introduction, we used it at this point in order to
obtain a first estimation of dynamic exponent $z$. However, we will
perform a critical analysis of the obtained results and subsequently we will
perform a second (more accurate) estimation of $z$. 
We carried out simulations up to $10^4$ MCS for segmentation steps $k=3,4,5$
and $6$ ($L=27,81,243$ and $729$, respectively) and determined the dynamic
exponent $z$ using an FSS analysis of the Binder's cumulant. In fact, right
at $T_c$, the dynamic exponent $z$ can be determined from the Binder's
cumulant according to the following scaling relation

\begin{equation}
U\left(t,L_1\right) =U(tb^{z},L_2) ,  \label{ec9}
\end{equation}

\noindent where $b=\frac{L_2}{L_1}$. Figure \ref{fig7} shows the data collapse
obtained when the time scale of the system of size L$_1$ size is rescaled by
a factor $\left(\frac{L_2} {L_1}\right)^z$. The results obtained by
rescaling lattices of sizes $81/27$, $243/81$, and $729/243$ are $z =
2.76(2), 2.65(1)$ and $2.60(3)$, respectively. So, we observed a systematic
decrease in $z$ when the segmentation step of the fractal is increased, and
consequently it is no longer valid to set a single value of the dynamic
exponent for all segmentation steps. This observed behavior is similar to an
observation reported previously \cite{7} where the fixed point intersection
of the Binder's cumulant for different sizes was replaced by a sequence of
intersection points occurring at ``effective'' critical temperatures, while
the actual critical temperature was defined as the limit for $%
k\rightarrow\infty$.

Summing up, the FSS method applied to the dynamic behavior of Binder's
cumulant only allow us to establish an upper bound to the dynamic exponent
given by $z = 2.60(3)$. 

In order to obtain an independent estimation of the dynamic exponent $z$
that doesn't involve calculations with segmentation steps smaller than 6 
we study the scaling behaviour of the time correlation function:

\begin{equation}
C(r,t)= \left[ \left\langle \frac 1N\sum_{i=1}^Ns_i(t)\,\,s_{i+r}(t)\right\rangle 
\right]
,  \label{ec91}
\end{equation}

\noindent where $i+r$ indicates a site displaced by $r$ lattice spacings
relative to site $i$. Our purpose is to study the onset of 
correlations between spins when an initially completely disordered 
system ($T=\infty$) has been quenched to $T=T_c$. Conventional critical scaling 
implies the following scaling form for $C(r,t)$:

\begin{equation}
C(r,t)= r^{-(d-2+\eta)} f_c(r/\xi(t)).  
\label{ec92}
\end{equation}
\noindent Assuming that the hyperscaling relation given by  
$d_{eff}= 2\beta /\nu + \gamma/\nu $ holds for this system, 
and using $\eta=2-\gamma/\nu $, we may replace 
$d-2+\eta$ in equation (\ref{ec92}) by $2\beta /\nu$. As $2\beta / \nu z$ 
has already been obtained directly in the simulations from the decay of the 
magnetization from the ordered state (equation (\ref{ec6})), and $\xi(t)$ is 
expected to behave as $t^{1/z}$, we may plot

\begin{equation}
r^{\left ( \frac{2 \beta}{\nu z} \right ) . z} C(r,t) ~~{\rm vs}~~ \frac{r}{t^{1/z}},  
\label{ec93}
\end{equation}

\noindent and look for the value of $z$ that make the curves to collapse.
This procedure has been applied for Humayun and Bray \cite{Bray}
to obtain $z$ for the Ising model for $d=2$.

\begin{figure}[ht]
\includegraphics[height=12cm,width=8cm,angle=-90]{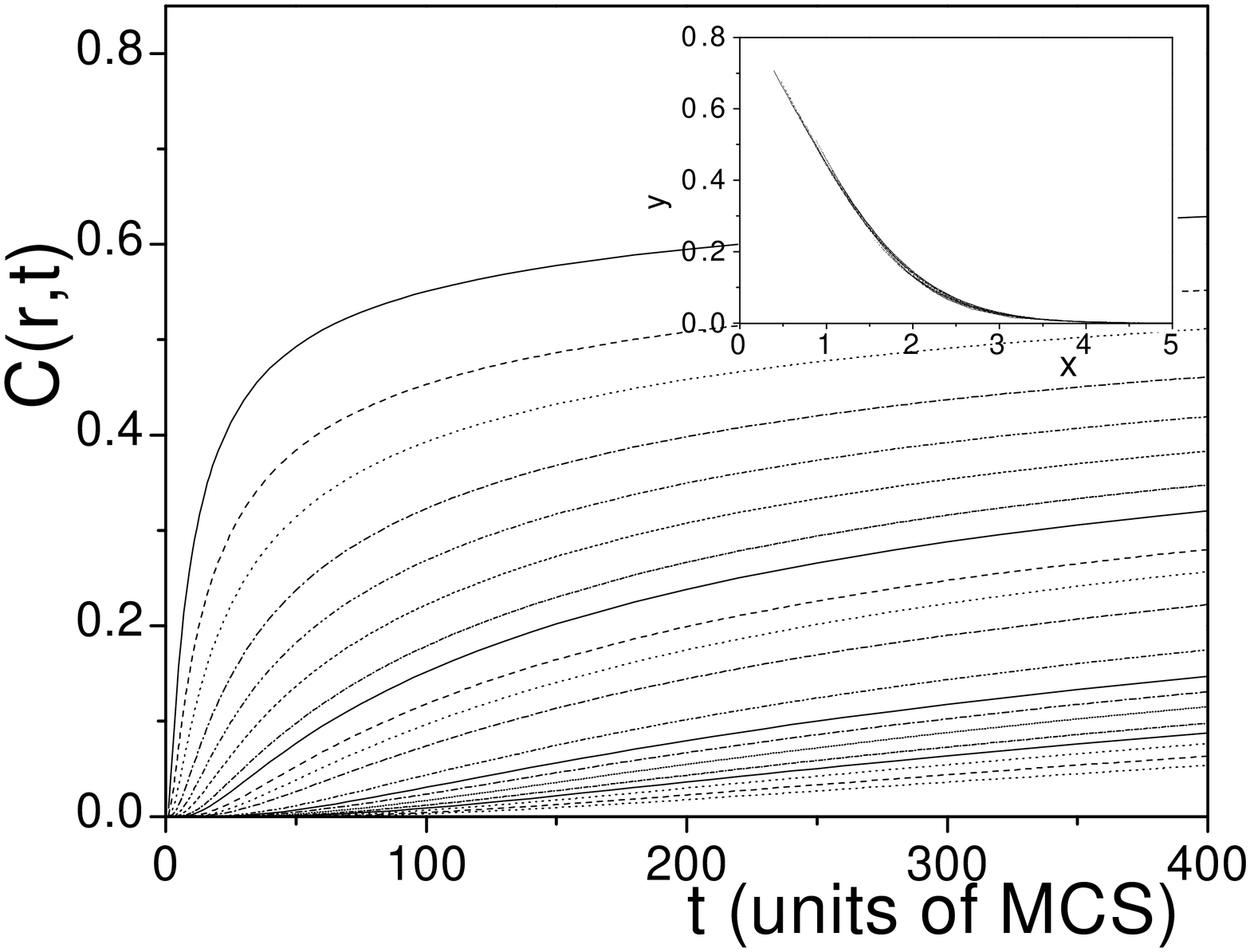} 
\vskip 1.0 true cm
\caption{Time correlation functions $C(r,t)$ for $r$ ranging from $r=\sqrt 17$
(upper curve) to $r = \sqrt 530$ (lower curve) obtained for the SC(3,1) 
and $k = 6$. The inset shows plots of the scaled correlation
$y = r^{\left ( \frac{2 \beta}{\nu z} \right ) . z} C(r,t)$
as a function of the scaled variable $x = r/t^{1/z}$,
for $z=2.55$ and $\frac{2 \beta}{\nu z}$ =0.0682. We have taken the value 
of $\frac{2 \beta}{\nu z}$ from the decay of $M(t)$ obtained in this 
work and adjusted the value of $z$ that gives the best collapse of the curves.
For $z=2.50$ and $z=2.60$ the curves show noticeable deviations
from the collapsed form (not shown here for the sake of space).}
\label{fig9}
\end{figure}

In figure \ref{fig9} we show plots of $C(r,t)$ as a function of $t$ 
obtained for different values of $r$ ranging from 4 to 23.
The inset shows the best collapse of the curves obtained for 
$z=2.55$. This value was obtained performing a fit of the scaled data to 
a 4-parameter function given by

\begin{equation}
f(x)= mx + b -m \pi^{-1/2}\int_0^x \int_{-\infty}^{ \frac{u-\alpha}{\sigma}}
    e^{-v^2} du dv,  
\label{ec94}
\end{equation}

\noindent that we have empirically found to fit the data quiet well.
We have also checked that other functional dependences for $f(x)$
that also fit the data only modifies the value of $z$ by less than 0.01.
The obtained value for $k = 6$, i.e.  $z =2.55(1)$ is consistent with the 
trend observed from the FSS analysis of Binder's cumulant and sets our upper 
bound for the dynamic exponent.

The relationship $d_{eff}/z=0.693$ has been determined quite accurately by
using three different kinds of measurements: the Binder's cumulant, the
slope of the second moment of the order parameter and the autocorrelation
function (see Table \ref{tab4}). So, taking  $z = 2.55$ the effective dimension 
becomes $d_{eff} \simeq 1.77$ , i.e. a figure that is noticeably smaller than the
Haussdorf dimension $d_{H} = 1.89$ of the SC(3,1). It should be noted that
previous estimations using FSS studies gave values of $d_{eff}$ very close
to $d_{H}$ (for a compilation of published results see table \ref{tab1}).
However, Pruessner {\it et. al} \cite{8}, have estimated $d_{eff} \simeq 1.7$%
, also smaller than the Haussdorf dimension.

On the other hand, using the value $1/\nu z = 0.282$ determined from the
slope of the logarithmic derivative of the order parameter (see 
equation (\ref{ec7}) and $z = 2.55$, our estimation for the lower bound of
the correlation length exponent becomes $\nu \simeq 1.39$, which is
significantly larger than the exact value of the Ising model given by $\nu =
1$.

\section{Discussion and conclusions.}

It is shown that the Short-Time Dynamics approach is a powerful method for
the study of the critical behavior of the Ising Model embedded in a fractal
structure, where the translational symmetry is changed for the scale
invariance. This method allows us to
obtain a self-consistent determination of the critical temperature and the
complete set of the critical exponents. This self-consistence is achieved by 
using three different initial conditions for the study of the dynamics.
 
The critical temperature determined in the present work for $k=6$ (an
upper bound for this system)
is in agreement with the value reported by Prussener et al. \cite{8}, which
was obtained from equilibrium measurements. We note that these authors have
used the Slope method, a procedure that is free of the finite size effects
involved in FSS calculations.
A critical discussion of the values reported by other authors employing different
techniques is presented in the discussion of the results listed in Tables 
\ref{tab1} and \ref{tab2}.

The exponent $\theta $ of the initial increase in $M(t)$ determined for the
segmentation step $k = 6$ and extrapolated to the $m_{0} \rightarrow 0$
limit is slightly smaller than the value corresponding to the two
dimensional Ising model (see Table \ref{tab1}). Our result is also in
agreement with that reported by Zheng et al. \cite{9}, obtained by locating
the spins at the vertices of the SC(3,1) and for the same segmentation step.
In this way, these results suggest that the fractal structure does not
significantly affect the exponent $\theta $.

The dynamic exponent $z$ has been obtained by means of two independent
measurements. Binder's cumulant method allow us to determine the 
decreasing trend of $z$ when $k$ is increased, so we obtained an 
upper bound given by $z = 2.60$. Further analysis of correlation 
functions allow us to improve the estimation of such an upper bound
which is given by $z = 2.55$.

Our value for the exponent $\beta = 0.121$ of the order parameter 
for the segmentation step $k = 6$, is also in agreement with the trend
of the results reported by Pruessner et al. \cite{8} for segmentation steps
$k = 4$ and $k = 5$ (see  Table II), suggesting that our estimation
can be taken as a lower bound. This value is only slightly smaller
than the exact exponent corresponding to the 2d Ising magnet, namely $\beta =
0.125$. So, we should not disregard the possibility that
for $k \rightarrow \infty$ the order parameter critical exponent 
may adopt the same value for both systems.    

On the other hand, our estimations of the exponents $\gamma = 2.22$ 
of the susceptibility and $\nu = 1.39$ of the correlation length, 
are significantly larger than those obtained for the 2d
Ising system, namely $\gamma = 1.75$ and $\nu = 1$, respectively.
Observing the trend reported by Pruessner et al. \cite{8} (see Table II) 
it is expected that our estimation for $\gamma$ could be taken as a 
lower bound. Also, $\nu$ may be taken as a lower bound because it is
evaluated from the measurement of $\nu z$ that involves our estimation 
of an upper bound for $z$. 

The relationship $d_{eff}/z = 0.693$ has been determined quite accurately,
then taking $z = 2.55$ we conclude that $d_{eff} \sim 1.77 < d_H$.

\begin{figure}[ht]
\includegraphics[height=14cm,width=10cm,angle=-90]{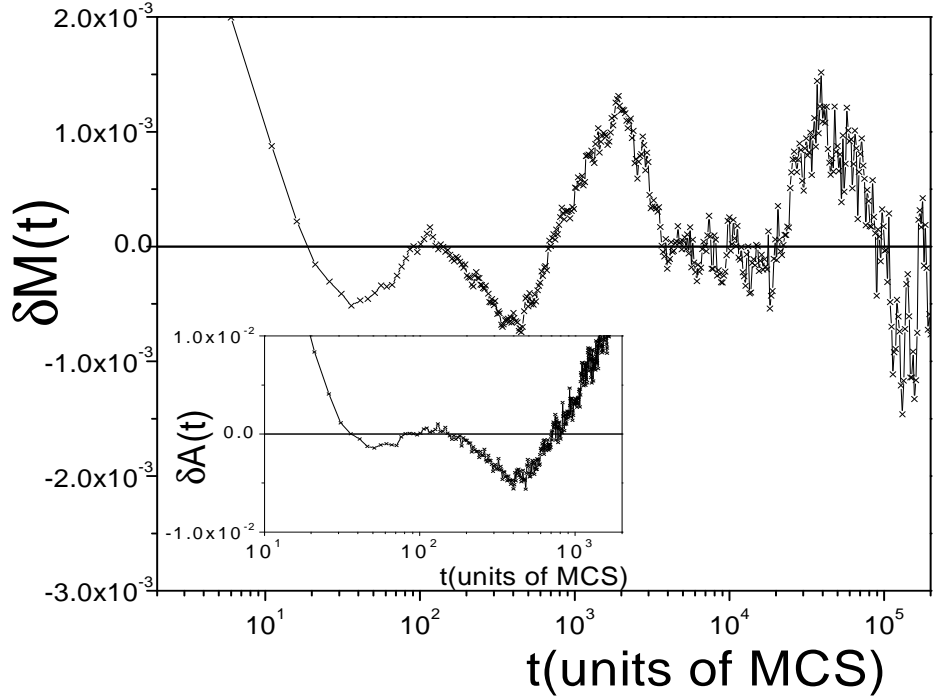} 
\vskip 1.0 true cm
\caption{Linear-log plots of the difference between the actual data 
corresponding to the decay of the order parameter obtained starting with
$m_{0} = 1 $ and the best power-law fit of the curve versus time.
The inset shows a similar plot but corresponding to the autocorrelation 
measured starting with $m_{0} = 0$.}
\label{fig8}
\end{figure}

Finally we would like to remark that due to the huge statistic achieved in
the evaluation of the dynamic properties, a soft oscillation around the
power-law decay of the magnetization was observed at criticality (see figure 
\ref{fig1}). The same oscillation was also observed for the disordered
initial state in the behavior of the autocorrelation (see figure \ref{fig6}). 
The oscillation can clearly be detected by subtracting the fitted power
law from the actual data, as shown in figure \ref{fig8}. This oscillation is
very nicely reproduced in both measurements 
(up to t $\simeq 2\times 10^3$ in figure \ref{fig8})
and, to the best of our knowledge this is the first evidence reported about
this interesting behavior of the dynamic properties of the Ising model in a
fractal substrate. We have clear signs that the observed oscillations are related
to the topological properties of the fractal lattice. A more
detailed investigation of these oscillations for the SC(3,1) and
other fractals will be published elsewhere \cite{osci}.

We would finally like to remark the self-consistency of our results 
obtained upon the application of the STD method to the study of the critical
behavior of the Ising model on fractal structures. However, it should be 
recognized that there are still discrepancies in the values of the 
critical exponents when STD-results are compared to those obtained using standard
finite-size scaling of equilibrium data. For example, in contrast to our data, 
recent FSS results of Carmona et al \cite{5} and Monceau et al. \cite{7} lead to 
an effective dimension almost equal to the Hausdorff one. So, we conclude that 
the origin of the discrepancies may be related to the fact that 
the critical behavior of the Ising magnet on fractal substrata is very particular, 
since it is linked to the dependence of most physical observable upon the number of 
iteration steps of the structure.

Acknowledgements: This work was supported by CONICET, UNLP, ANPCyT and
Fundaci\'on Antorchas (ARGENTINA). The A. von Humboldt Foundation (Germany)
is greatly acknowledged for the provision of valuable computer equipment.
The authors thanks Silvio Franz for fruitful discussions.

\end{document}